\documentclass[dvips]{acta}
\usepackage{supertabular,lscape,epsfig}
\usepackage{amssymb}
\usepackage{amsmath}
\SetPages{1}{5}
\SetVol{70}{2020}

\begin{document}

\begin{Titlepage}

\Title { On the Periods and Nature of Superhumps }

\Author {J.~~S m a k}
{N. Copernicus Astronomical Center, Polish Academy of Sciences,\\
Bartycka 18, 00-716 Warsaw, Poland\\
e-mail: jis@camk.edu.pl }

\Received{  }

\end{Titlepage}

\Abstract {It is commonly accepted that the periods of superhumps can be 
satisfactorily explained within a model involving apsidal motion of the 
accretion disk provided the frequency of the apsidal motion in addition to 
the dynamical term includes also the pressure effects. 
Using a larger sample of systems with reliable mass ratios it is shown,  
however, that this view is not true and the model requires further modifications. 
}
{accretion, accretion disks -- binaries: cataclysmic variables, stars: dwarf novae }

\section { Introduction } 

Superhumps are periodic light variations, with periods slightly longer than the 
orbital period, observed in dwarf novae during their superoubursts and in 
nova-like cataclysmic variables -- the so-called permanent superhumpers. 

There are two, competing models for superhumps: the tidal-resonance model and 
the irradiation modulated mass transfer model. 
According to the tidal-resonance (TR) model, first proposed by Whitehurst (1988) 
and Hirose and Osaki (1990), they are due to periodic enhancement of tidal stresses 
in an eccentric accretion disk undergoing apsidal motion (often incorrectly called 
"precession"). An important ingredient of this model is the 3:1 resonance between 
the orbital frequency of the binary system and the orbital frequency of the outer 
parts of the disk which is essential for the disk to become eccentric. 
This model, however, fails to explain many crucial facts (cf. Smak 2017 and references 
therein) but in spite of that is commonly accepted. 
The irradiation modulated mass transfer (IMMT) model (Smak 2009,2017), based on purely 
observational evidence, explains superhumps as being due to the periodically 
variable dissipation of the kinetic energy of the stream resulting from variations 
in the mass transfer rate which are produced by the modulated irradiation of the 
secondary component. 

The periods of superhumps and their interpretation have been the subject of numerous 
investigations. According to the most recent ones (Murray 2000, Montgomery 2001, 
Pearson 2006, Smith {\it et al.} 2007) they cannot be explained within 
purely dynamical theory but satisfactory agreement between theoretical predictions 
and observations is obtained when pressure effects are included. 

The main purpose of the present paper is to verify this conclusion by using a much 
larger sample of superhumpers with reliably determined mass ratios.


\section { The Data } 

The data set to be analyzed below consists of 26 cataclysmic variables, 
with independently determined mass ratios, taken from the recent compilation 
by McAllister {\it et al.} (2019, Tables 2 and B2) and supplemented from 
the compilation by Smith {\it et al.} (2007, Table 5).  
Included in our sample are also two helium CV's: AM CVn with $q=0.18\pm0.01$ 
(Roelofs {\it et al.} 2006) and YZ LMi with $q=0.041\pm 0.002$ 
(Copperwheat {\it et al.} 2011). 

It is worth noting that the number of objects with $P_{orb}<3$ hours in our 
sample (21) is much larger than in previous investigations (e.g. Pearson 2006, 
Table 3, or Smith {\it et al.} 2007, Table 3). On the other hand, however, 
the number of objects with $P_{orb}>3$ hours (5) remains very small. 
This is due to the fact that most of them belong to the SW Sex type 
(cf. Dhillon {\it et al} 2013) showing effects of the stream overflow which 
- regretfully - make determination of the mass ratio from radial velocities 
or from hot spot eclipses difficult and/or unreliable.

\section { The Basic Equations and Relations } 

We begin by listing equations and relations which will be used in further analysis. 
The superhump period $P_{SH}$ and the period of apsidal motion $P_{aps}$ 
are related by 

\beq
{1\over {P_{aps}}}~=~{1\over {P_{orb}}}~-~{1\over {P_{SH}}}~,
\eeq 

\noindent
or, in terms of the corresponding frequencies $\omega=2\pi/P$, 

\beq
\omega_{aps}~=~\omega_{orb}~-~\omega_{SH}~.
\eeq 

\noindent
The superhump period excess defined as

\beq
\epsilon_{SH}~=~{{P_{SH}-P_{orb}}\over {P_{orb}}}~,
\eeq 

\noindent
is related to the apsidal frequency by 

\beq
\epsilon_{SH}~=~{ {\omega_{aps}}\over {\omega_{orb}-\omega_{aps}} }~.
\eeq 

It is commonly assumed (e.g. Pearson 2006, Smith {\it et al.} 2007 and references 
therein) that the apsidal frequency is the sum of the dynamical and pressure terms 

\beq
\omega_{aps}~=~\omega_{dyn}~+~\omega_{press}~.
\eeq 

The ratio of the dynamical part of the apsidal frequency to the orbital frequency 
as a function of the mass ratio and the {\it effective radius} of the disk  
is given by (Hirose and Osaki 1990, Pearson 2006, Eqs.6 and 7) 

\beq
{{\omega_{dyn}}\over {\omega_{orb}}}~=~{3\over 4}~{q\over {(1+q)^{1/2}}}~r^{3/2}~
\sum_{n=1}^\infty~a_n~r^{2(n-1)}~,
\eeq

\noindent
where

\beq
a_n~=~{2\over 3}~(2n)~(2n+1)~\prod_{m=1}^n~\left({{2m-1}\over {2m}}\right)^2~.
\eeq 

The {\it effective radius} of the disk is commonly assumed to be equal to 
the 3:1 resonance radius which is given by 

\beq
r_{3:1}~=~{1\over {3^{2/3}(1+q)^{1/3} } }~.
\eeq

\section { The Pressure Term ? } 

Following earlier papers (Murray 2000, Montgomery 2001, Pearson 2006, 
Smith {\it et al.} 2007) we now determine the residuals $\Delta\omega$ 
between the observed apsidal frequency $\omega_{aps}$ and the dynamical 
term $\omega_{dyn}$ which, according to earlier authors, are expected 
to represent the pressure term $\omega_{press}$. 

First, using the observed values of $P_{orb}$, $\epsilon_{SH}$ and $q=M_2/M_1$
we determine the observed apsidal frequency $\omega_{aps}$ (Eq.4). 
Then, assuming -- as is commonly done --  that the {\it effective radius} 
of the disk is equal to the 3:1 resonance radius  
\footnote 
{replacing $r_{3:1}$ with $r_{tid}$ for systems with orbital periods above 
the period gap changes the results only slightly. } 
(Eq.8), we calculate the dynamical contribution $\omega_{dyn}$ (Eqs.6 and 7) and, 
finally, the difference: $\Delta\omega=\omega_{aps}-\omega_{dyn}$.  

\begin{figure}[htb]
\epsfysize=10.0cm 
\hspace{2.0cm}
\epsfbox{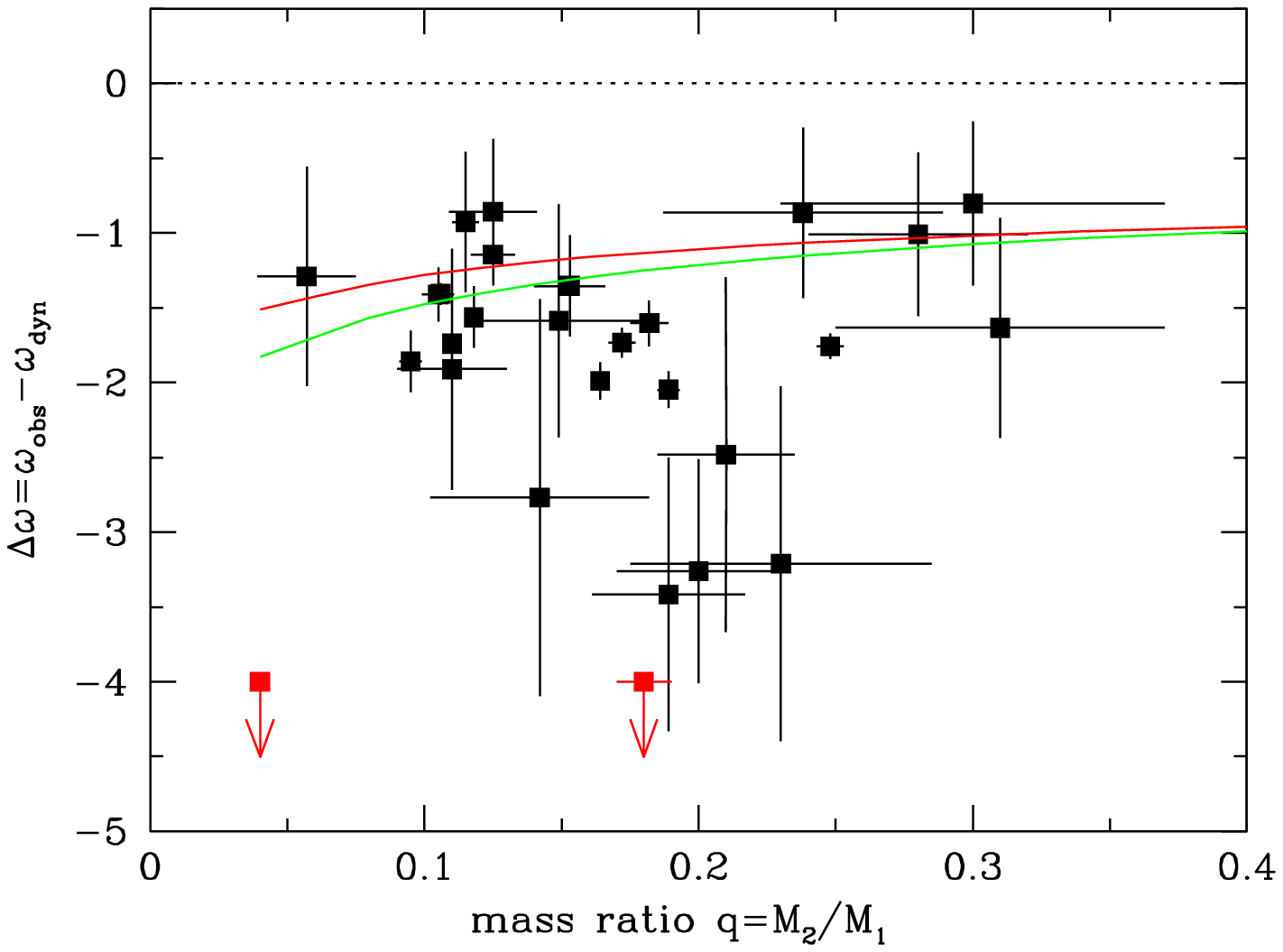} 
\vskip -3.5truecm
\FigCap {The residuals $\Delta\omega$ (see text for details) are plotted against 
the mass ratio. Red symbols represent the two helium CV's. Red and green lines 
are theoretical $\omega_{press}=f(q)$ relations from, respectively, 
Montgomery (2001) and Pearson (2006). 
}
\end{figure}

Results, presented in Fig.1, can be summarized as folows: 

{\parskip=0truept {
(1) For systems with $P_{orb}<3$ hours the residuals $\Delta\omega$ show large 
scatter and/or appear to be correlated with the mass ratio. 

(2) Systems with $P_{orb}>3$ hours form a separate group. 

(3) The residuals for helium CV's are very large: 
$\Delta\omega=-5.0\pm0.3$ for YZ LMi and $\Delta\omega=-21.9\pm0.9$(!) for AM CVn;  
\footnote 
{the problem with the AM CVn systems was already noted by Pearson (2007). }

(4) The theoretical $\omega_{press}=f(q)$ relations (Montgomery 2001, Pearson (2006) 
fail to represent the real data. 
}}
\parskip=12truept

\section { Discussion } 

Until now, as mentioned in the Introduction, it has been believed that 
the periods of superhumps can be satisfactorily explained when the apsidal 
frequency is assumed to be the sum of the dynamical and pressure terms. 
Results presented above imply that this is not true. Therefore the basic model 
involving the apsidal motion of the disk requires substantial modifications. 

\begin{figure}[htb]
\epsfysize=10.0cm 
\hspace{2.0cm}
\epsfbox{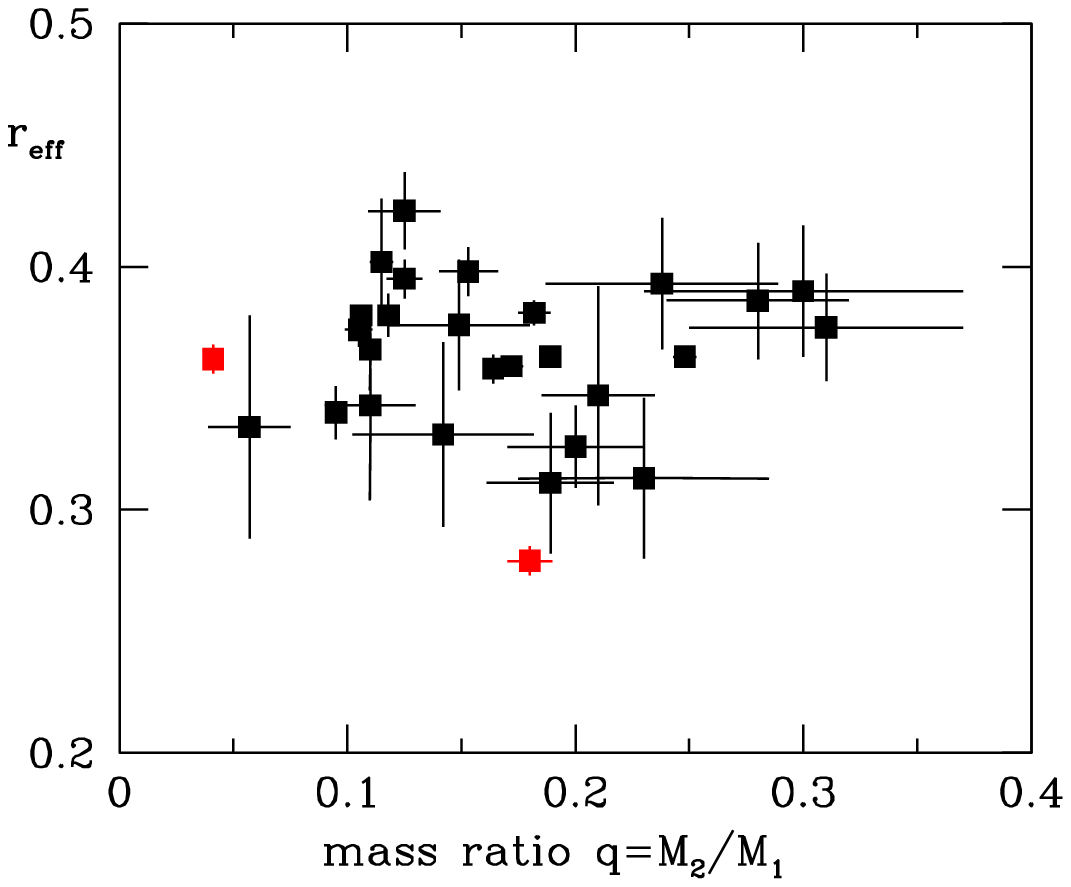} 
\vskip -4.5truecm
\FigCap {The {\it effective radii} $r_{eff}$ (see text for details) are plotted 
against the mass ratio. Red symbols represent the two helium CV's. 
}
\end{figure}

Searching for possible clues we follow Pearson (2006) and determine the 
{\it effective radii} $r_{eff}$ at which the apsidal frequency calculated using 
only the dynamical term (Eqs.6 and 7) would be equal to the observed frequency. 
Results, presented in Fig.2, are remarkable: they show that {\it all} values of 
$r_{eff}$, including those representing the two AM CVn systems, fall between 
$\approx 0.3$ and $\approx 0.4$. The significance of this results is, however, 
not yet clear. 

\begin {references} 

\refitem {Copperwheat, C.M. {\it et al.} } {2011} {\MNRAS} {410} {1113}

\refitem {Dhillon, V.S., Smith, D.A., and Marsh, T.R.} {2013} {\MNRAS} {428} {3559} 

\refitem {Hirose, M., and Osaki, Y.} {1990} {\it Publ.Astr.Soc.Japan} {42} {135}

\refitem {McAllister, M. {\it et al.} } {2019} {\MNRAS} {486} {5535}

\refitem {Montgomery, M.M.} {2001} {\MNRAS} {325} {761}

\refitem {Murray, J.R.} {2000} {\MNRAS} {314} {L1}

\refitem {Pearson, K.J.} {2006} {\MNRAS} {371} {235} 

\refitem {Pearson, K.J.} {2007} {\MNRAS} {379} {183} 

\refitem {Roelofs, G.H.A., Groot, P.J., Nelemans, G., Marsh, T.R., Steeghs, D.} 
         {2006} {\MNRAS} {371} {1231} 

\refitem {Smak, J.} {2009} {\Acta} {59} {121} 

\refitem {Smak, J.} {2017} {\Acta} {67} {273} 

\refitem {Smith, A.J., Haswell, C.A., Murray, J.R., Truss, M.R., and Foulkes, S.B.} 
         {2007} {\MNRAS} {378} {785} 

\refitem {Whitehurst, R.} {1988} {\MNRAS} {232} {35} 

\end {references}

\end{document}